\documentclass[journal=jacsat,manuscript=article]{achemso}

\usepackage[version=3]{mhchem} 
\usepackage{xcolor}
\usepackage[utf8]{inputenc}
\usepackage{siunitx}
\usepackage{booktabs}

 \usepackage{mciteplus}

\usepackage{booktabs}
\usepackage{array}
\usepackage{graphicx}

\author{Manobina Karmakar}
\affiliation[LMU]
{Chair in Hybrid Nanosystems, Nanoinstitute Munich, Faculty of Physics, Ludwig-Maximilians-Universität München, Königinstraße 10, 80539 München, Germany}
\author{Pavel Klok}
\affiliation[BUT]
{Brno University of Technology, Faculty of Mechanical Engineering, Institute of Physical Engineering, Technická 2, 61669 Brno, Czech Republic}

\author{Filip Ligmajer}
\affiliation[BUT]
{Brno University of Technology, Faculty of Mechanical Engineering, Institute of Physical Engineering, Technická 2, 61669 Brno, Czech Republic}
\alsoaffiliation[CEITEC]
{Brno University of Technology, Central European Institute of Technology, Purkyňova 123, 61200 Brno, Czech Republic}
\email{filip.ligmajer@vutbr.cz}
\author{Leonardo de S. Menezes}
\affiliation[LMU]
{Chair in Hybrid Nanosystems, Nanoinstitute Munich, Faculty of Physics, Ludwig-Maximilians-Universität München, Königinstraße 10, 80539 München, Germany}
\alsoaffiliation[UFPE]
{Departamento de F\'isica, Universidade Federal de Pernambuco, 50670-901 Recife-PE, Brazil}
\email{L.Menezes@physik.uni-muenchen.de}
\title[An \textsf{achemso} demo]
  {Tunable hyperbolic metamaterials for brightening single-photon emission}

\begin{document}




\begin{abstract}
Hyperbolic metamaterials (HMMs) are widely used to enhance spontaneous emission through their exceptionally large photonic density of states, yet large Purcell factors ($>$100) rarely translate into brighter emission, exposing both a practical limitation of HMMs as brightening platforms and a flaw in the Purcell factor itself as a figure of merit. Here we systematically evaluate spontaneous emission enhancement and photon outcoupling in metal–dielectric HMMs (TiO$_2$/Ag and phase-change Sb$_2$S$_3$/Ag) using effective medium theory, transfer matrix calculations, and full-wave FDTD simulations. We show that planar HMMs with Purcell factors exceeding 100 have external quantum efficiencies below 0.01, yielding an actual radiative Purcell factor ($\text{PF}_\text{rad}$) of only 1–2. Nanopatterning the HMM into a photonic crystal efficiently couples trapped high-$k$ modes into free space, raising $\text{PF}_\text{rad}$ by an order of magnitude. Incorporating the phase-change material Sb$_2$S$_3$ further enables dynamic tuning of the hyperbolic spectral window and active control over emission brightness: an optimized nanostructured design reaches $\text{PF}_\text{rad}$ $\sim$ 10–15 in the near-infrared, with switching contrast exceeding 15× upon phase transition. These results establish the radiative Purcell factor as a more reliable figure of merit than the conventional Purcell factor for engineering quantum emitters, and provide a practical route toward actively tunable, bright single-photon sources based on hyperbolic metamaterials.
\end{abstract}

\section{Introduction}
Solid-state single-photon emitters (SPEs) are essential building blocks for scalable photonic quantum information processing \cite{OBrien2007OpticalQC,doi:10.1021/acs.jpclett.2c03674}. Despite deterministic single-photon generation, currently available SPEs face several challenges such as low emission rate, fast decoherence, and limited external control that hinder practical deployment\cite{Lounis_2005}. Despite deterministic single-photon generation, currently available SPEs still face several challenges, such as low emission rate, fast decoherence, and limited external control, that hinder practical deployment\cite{Lounis_2005}. A general strategy to address these challenges is to embed SPEs in nanostructured photonic environments with elevated photonic density of states (PDOS), thereby enhancing light-matter interactions\cite{https://doi.org/10.1002/adom.202202759} and enabling brighter single-photon emission. Engineered optical cavities, like plasmonic nanostructures\cite{Akselrod2014}, Bragg reflectors\cite{doi:10.1021/acsphotonics.4c01873, Somaschi2016, PhysRevLett.122.113602}, waveguides\cite{Gusken2023}, and various photonic crystal structures\cite{PhysRevLett.113.093603} allow enhanced light-matter interaction and Purcell enhancement. Amongst these,  hyperbolic metamaterials (HMMs), a family of engineered uniaxial, anisotropic materials that display metallic and dielectric properties in perpendicular directions, are of particular interest. HMMs offer an extremely high PDOS, governed by the open topology of the isofrequency surface, resulting in extremely high Purcell factors (PFs) \cite{Shekhar2014,Poddubny2013}.

The spontaneous emission rate enhancement provided by a photonic environment is conventionally quantified by the PF, defined as the ratio of the emitter's decay rate in the cavity $\Gamma$ to its decay rate in vacuum $\Gamma_0$: $\text{PF} = \Gamma/\Gamma_0 = (\Gamma_\text{rad}+\Gamma_\text{non-rad})/\Gamma_0$. Although widely used, the PF captures the total decay rate, including both radiative and non-radiative contributions, rather than the useful radiative emission that can actually be collected experimentally. In plasmonic and hyperbolic structures in particular, a substantial fraction of the enhanced decay is funneled into non-radiative channels, such as Ohmic losses and lossy surface or bulk plasmon modes\cite{Barnes01041998,Li:16}. As a result, a large PF does not necessarily correspond to a brighter emitter or improved photon extraction. This distinction is especially important for quantum photonic applications, where device performance is ultimately determined by the number of usable photons rather than by the total decay rate. Although a few studies highlighted the insufficiency of the figure of merit\cite{Akselrod2014, Koenderink:10}, a single, reliable metric is still lacking.

In 2012, Krishnamoorthy et al. first predicted that the extreme anisotropy of HMMs could support significantly enhanced spontaneous emission rates, positioning these materials as promising platforms for brightening quantum emitters \cite{doi:10.1126/science.1219171}. This study motivated a new line of research aimed at brightening quantum emitters\cite{doi:10.1021/acsphotonics.7b00767, doi:10.1021/acsanm.0c02186}. Researchers observed PFs at least an order of magnitude higher than those of comparable contemporary photonic structures\cite{PhysRevB.92.195127,oea-2021-0031-Andrei}. 
However, despite these large predicted and measured PFs, experimentally observed emitter efficiency enhancements have remained modest,\cite{oea-2021-0031-Andrei} typically below a factor of 10, leaving unresolved questions about whether the predicted Purcell enhancement in HMMs is actually extractable as usable photons. 
Despite the promise of high Purcell enhancement, the suitability of these metamaterials as emitter-brightening platforms remains inconclusive, and the suitability of PF as the figure of merit is also still questionable. Moreover, quantum networks and on-chip devices essentially require active modulation of their emission rates to securely process quantum information\cite{Luo2023, https://doi.org/10.1515/nanoph-2024-0550, Lounis_2005}. Passive photonic structures lack the dynamic reconfigurability needed for real-time control of spontaneous emission.


Motivated by these inconsistencies between reported high PFs and modest photonic performance, our work has two central aims: first, to introduce a more reliable figure of merit for evaluating the efficiency of photon-emission-brightening platforms; and second, to investigate the potential of HMMs for external tunability, enabling active modulation for quantum photonic applications. To address the \textit{first aim}, we redefine the relevant figure of merit as the radiative PF and use it to re-evaluate HMM platforms, correcting for the ambiguities introduced by the conventional PF Using a combination of effective medium theory, transfer matrix calculations, and full-wave finite-difference time-domain (FDTD) simulations, we systematically compare conventional Purcell enhancement with the radiative PF in planar and nanostructured HMM architectures. We demonstrate improved outcoupling of a well-studied HMM metal-dielectric stack of TiO$_2/$Ag through photonic crystal (PhC) patterning over the stacks. We analyze what relative emitter position is optimal for the highest radiative PF, showing that HMM PhCs offer the highest outcoupling when the emitter is closest to the surface. The optimized HMM PhC achieves radiative PFs ($\text{PF}_\text{rad}$) exceeding 10, representing an order-of-magnitude enhancement over the comparatively low values ($\sim$1--2) obtained for unpatterned one-dimensional HMMs. 

To address the \textit{second aim}, we investigate how incorporation of a phase-change material (Sb$_2$S$_3$) into the HMM design can lead to a dynamically reconfigurable platform in which both the hyperbolic spectral window and the emission brightness can be actively tuned through reversible phase transitions. For an emitter placed near an optimized Sb$_2$S$_3$/Ag HMM PhC, we predict $\text{PF}_\text{rad}$ of approximately 10--15 in one phase and 1--8 in the other, corresponding to a brightness switching contrast exceeding an order of magnitude ($>15$) in the technologically important near-infrared spectral region. Taken together, this work provides both a revised framework for evaluating emission-enhancing photonic structures and a practical route toward tunable platforms for high-brightness SPEs.

\section{Results and discussion}

\begin{figure*}[!b]
\begin{center}
\includegraphics[]{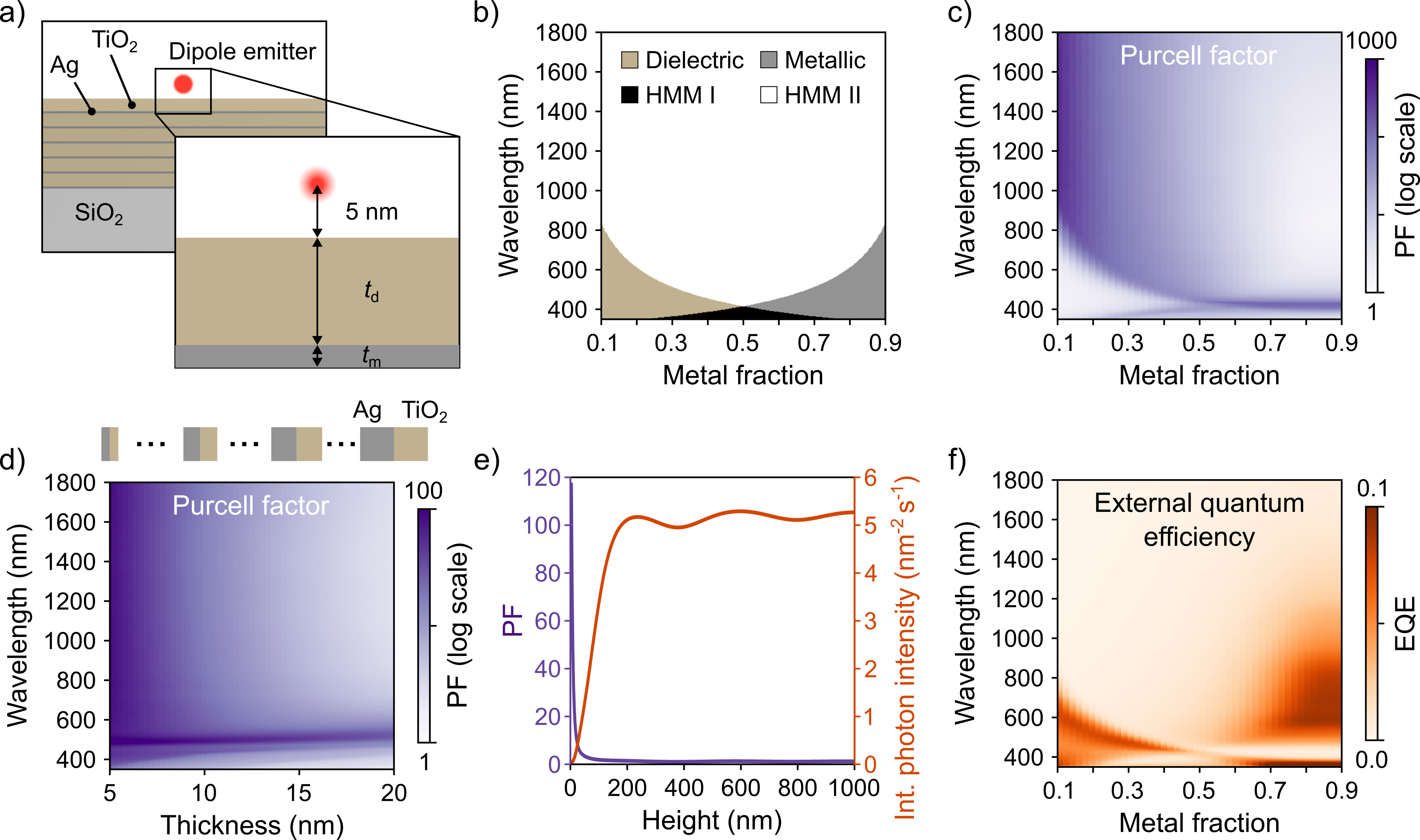}\caption{(a) Schematic of the simulation setup: a horizontal dipole emitter is placed 5\,nm above a TiO$_2/$Ag stack with 6 pairs of alternating layers on top of a SiO$_2$ substrate. (b) Effective medium theory predicts the spectral windows for dielectric, metallic, and hyperbolic (type I and type II) behaviors. (c) PF map of the HMM stack as a function of metal fraction and wavelength. Dielectric layer thickness ($t_\text{d}$) has been fixed to 10\,nm. (d) PF as a function of metal layer thickness with a constant $f=0.5$. (e) Variation of the integrated photon intensity and PF as a function of emitter-to-surface distance for the TiO$_2/$Ag stack corresponding to $f = 0.2$ at 800\,nm. (f) External quantum efficiency as a function of metal fraction and wavelength. 
\label{fig1}}
\end{center}
\end{figure*}

HMMs can be realized by periodically stacking alternating layers of metallic and dielectric materials\cite{Lee2022}, as illustrated in Fig.~\ref{fig1}(a). This configuration creates the necessary anisotropy of dielectric functions, in-plane ($\epsilon_{||}$) and out-of-plane ($\epsilon_{\perp}$). Specifically, hyperbolicity of type I corresponds to $\epsilon_{||}> 0$ and $\epsilon_{\perp}< 0$, while hyperbolicity of type II corresponds to $\epsilon_{||}< 0$ and $\epsilon_{\perp}> 0$\cite{Li:16}. The specific hyperbolic character of a spectral region is governed by the metal fraction $f=t_\text{m}/(t_\text{m}+t_\text{d})$, where $t_\text{m}$ ($t_\text{d}$) is the thickness of the metal (dielectric) layer. We utilize effective medium theory (EMT) \cite{Markel:16} to identify the effective dielectric, metallic, and hyperbolic regions of a system of Ag and TiO$_2$ and display the results in Fig.~\ref{fig1}(b). A broad spectral range from visible to the near infrared exhibits type II hyperbolicity, while there is also a type I region in the ultraviolet range around a metal fraction of \SI{0.5}. The PFs provided by TiO$_2/$Ag stacks were calculated as a function of metal fraction and wavelength through analytical Green's function formalism and the transfer matrix method\cite{novotny2012principles}. We assume a horizontally-polarized dipolar quantum emitter located \SI{5}{\nm} above the HMM surface. We restrict our analysis to horizontally-polarized dipoles, which couple most efficiently to the in-plane high-$k$ modes responsible for hyperbolic dispersion~\cite{10.1021/acsphotonics.0c01219} and are less susceptible to non-radiative quenching near metallic layers. In practice, emitters such as quantum dots or color centers may have arbitrary or randomly oriented transition dipole moments, so the reported $\text{PF}$ values will likely represent an upper bound for a given emitter ensemble. Note that to minimize emitter quenching, the top surface of the HMM stack is always the dielectric material. 
The resulting PF values in the type-II hyperbolic region, as can be identified from Fig.~\ref{fig1}(b), exceed 100, with the highest values occurring at low metal fractions (Fig.~\ref{fig1}(c)). The enhancement of PF is maximized when the unit-cell thickness of the stack is minimized, allowing a higher cut-off for the allowed $k$ vectors \cite {Poddubny2013}. Fig.~\ref{fig1}(d) confirms this effect as we plot the PF at $f=0.5$ as a function of the single layer's thickness. 

While the high PFs naturally suggest the suitability of the hyperbolic region for increasing spontaneous emission rates, one must also evaluate photon outcoupling to verify its applicability as a photon-brightening platform. A well-known figure of merit for this effect, external quantum efficiency (EQE), quantifies the fraction of radiative emission and is defined as the ratio of total radiative power ($P_r$) to total emitted power ($P_r+P_{nr}$; the sum of radiative and non-radiative powers). To estimate the experimentally collectible radiative output, we calculate EQE by integrating $P_r$ over the upper hemisphere only, essentially excluding the modes propagating into the substrate (i.e., EQE = $\frac{P_r^U}{P_r+P_{nr}}$; $U$ superscript indicates the upper hemisphere). In Fig.~\ref{fig1}(f), we show the calculated EQE of the TiO$_2/$Ag stack as a function of the metal fraction and wavelength. It displays low values ($<0.01$) in the hyperbolic region and somewhat higher values ($>0.07$) in the effective metallic region, confirming the experimental results highlighted in the Introduction. Such a low EQE is a classic manifestation of photon outcoupling quenching near a metallic (or dielectric) mirror\cite{DREXHAGE1974163}. To investigate this effect in greater detail, we plot both the photon intensity integrated over the upper hemisphere (i.e., away from the stack) and the PF as functions of the distance between the dipole emitter and the HMM surface (Fig.~\ref{fig1}(e)). While the suppression of photon outcoupling at small distances and its recovery beyond approximately 100\,nm are clearly evident, the PF decreases to very low values at these larger distances. In other words, although photon extraction is improved at the optimized emitter–surface separation, the influence of the hyperbolic modes becomes negligible, suggesting that hyperbolicity contributes only weakly to the observed emission enhancement under these conditions.

A comparison of Fig.~\ref{fig1}(c) and \ref{fig1}(f) clearly indicates that hyperbolicity correlates with lower photon collection. We note that the 5\,nm separation of the dipole from the metasurface affects photon outcoupling, as it is quenched by non-radiative and Ohmic losses in the metal\cite{DREXHAGE1974163,PhysRevB.95.155424}. In fact, the photon collection from an effective hyperbolic region is weaker than that from the metallic region, indicating very strong non-radiative loss channels that contribute to high PFs. A few earlier studies \cite{https://doi.org/10.1002/adom.202000368, PhysRevB.95.155424} have highlighted this issue. The comparison between the PF and the EQE demonstrates that neither quantity alone adequately characterizes the performance of an emission-enhancing platform. While the PF quantifies the total decay enhancement, the EQE, on the other hand, ignores the increase in the spontaneous-emission rate. This observation motivates us to suggest the use of the radiative PF $\text{PF}_\text{rad}$ introduced below, which combines both effects into a single experimentally relevant metric. 

$\text{PF}_\text{rad}$ is defined as the ratio of the radiated power of an emitter in an optical medium to that in vacuum.\cite{Krasnok2015} Unlike the conventional PF, which includes both radiative and non-radiative decay channels, $\text{PF}_\text{rad}$ quantifies only the power emitted into radiative modes that can, in principle, be collected. In other words, $\text{PF}_\text{rad}$ directly measures the brightness enhancement experienced by an emitter and, therefore, provides a more meaningful figure of merit for evaluating photon-brightening platforms. Since this quantity excludes energy dissipated through absorption and other non-radiative processes, it naturally combines the effects of spontaneous-emission enhancement and photon outcoupling into a single experimentally relevant quantity. To estimate $\text{PF}_\text{rad}$, we run FDTD simulations of the various platforms discussed so far (see Methods for more details). The results are summarized in Table~\ref{tab:photonic}, where we compare the PF, EQE, and $\text{PF}_\text{rad}$ values for different combinations of substrates, mirrors, HMM stacks, and dipole-to-surface distances. These results confirm that bare dielectric substrates provide only small emission brightening, while metallic mirrors offer larger values, provided the emitter is not too close to the surface. In other words, both PF and EQE must be significant for emission brightening ($\text{PF}_\text{rad} >$ 1). TiO$_2/$Ag HMM offers a PF that is an order of magnitude higher than that of the Ag mirror, but their values of $\text{PF}_\text{rad}$ are comparable and do not justify the increased fabrication complexity associated with HMMs. To ensure efficient photon outcoupling from the HMM and thus increase its EQE, we decided to investigate nanohole etching into the HMM stack. We expected that by sculpting the HMM into the form of a photonic crystal, we could break the in-plane translational symmetry, facilitate the outcoupling of high-$k$ waves that would otherwise be trapped inside the HMM, and thus achieve an enhancement of $\text{PF}_\text{rad}$ beyond the current state of the art. 

\begin{table}[tb]
\centering
\caption{Comparison of different metrics for brightness enhancement}
\label{tab:photonic}
\resizebox{\textwidth}{!}{
\begin{tabular}{cccccc}
\toprule
\textbf{Photonic}& \textbf{Emitter-to-surface} & \textbf{Wavelength} & \textbf{PF} & \textbf{EQE (\%)} & $\mathbf{PF}_\mathbf{rad}$\\
 \textbf{platform}& \textbf{distance (nm)} & \textbf{range (nm)} & & & \\\midrule

Ag mirror & 5 & 400 -- 1600 & 20 -- 300 & 0.1 -- 0.9  &  0.02 -- 0.7 \\

Ag mirror & 150 & 400 -- 1600 & 0.4 -- 1 & \>90 & 1--4 \\

SiO$_2$ substrate & 5 & 400 -- 1600 & $\sim$ 1.3 & $\sim$ 21  & $\sim$ 0.7 \\

SiO$_2$ substrate & 150 & 400 -- 1600 & $\sim$ 1.0 & 30 -- 60  & 0.9 -- 1.5 \\

TiO$_2$/Ag 1D (f = 0.2) HMM & 5 & 500 -- 1800 & 1.5 -- 200 & $<$ 10  & 1--2 \\

TiO$_2$/Ag 1D ($f = 0.2$) HMM & 150 & 500 -- 1800 & 0.35 -- 1.5 & 20 -- 75 & 0.7 -- 3.3 \\
\bottomrule
\end{tabular}}
\end{table}

\begin{figure*}[!h]
\begin{center}
\includegraphics[width=1.0\linewidth]{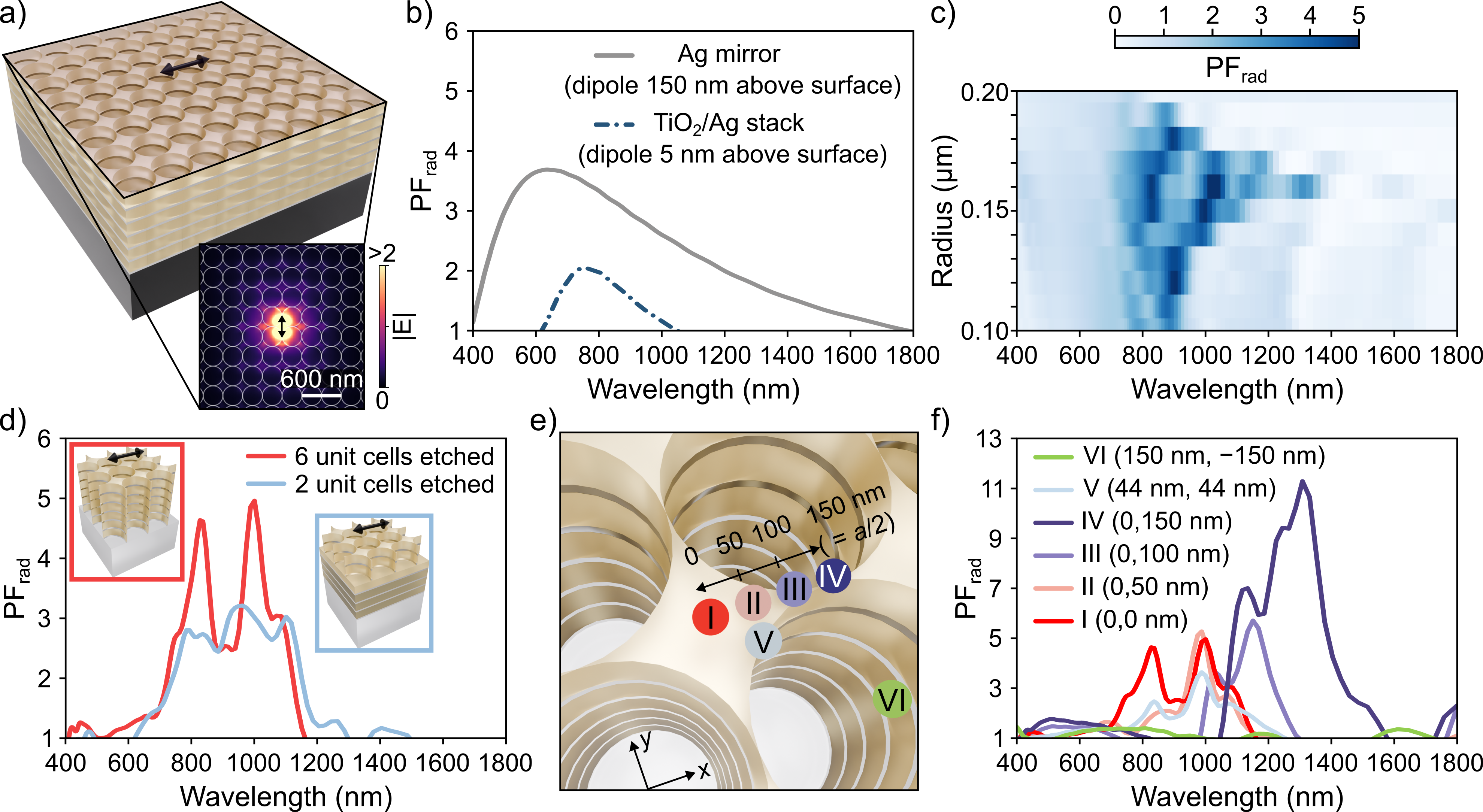}\caption{Photon emission efficiency and collection enhancement in a 3D structured HMM: (a) Schematic of a 3D photonic crystal (PhC) based on an etched HMM. An in-plane polarized electric dipole emitter is represented by a black arrow. Inset: Electric field amplitude $|E|$ at the surface of HMM PhC at wavelength of \SI{800}{\nm}, corresponding to periodicity $a = $ \SI{0.3}{\micro\meter} and holes' diameter \SI{0.15}{\micro\meter}. The dipole polarization is indicated by the black arrow. The corresponding spectral location in (c) is marked with a green diamond. (b) Reference $\text{PF}_\text{rad}$ of two systems: a horizontal dipole placed \SI{150}{\nm} above an Ag mirror and the same dipole \SI{5}{\nm} above a TiO$_2/$Ag HMM stack. (c) Comparison of $\text{PF}_\text{rad}$ of various HMM PhCs of periodicity a = 0.3\,$\mu$m and varying hole diameters (dipole \SI{5}{\nm} above the surface) (d) Comparison between the $\text{PF}_\text{rad}$ of a structured PhC with only two HMM layers etched and with full six HMM layers etched. (e) Schematic of six in-plane dipole positions relative to the PhC unit cell, labeled by Roman numbers and respective colors. (f) Comparison of PF$_{\text{rad}}$ for a dipole placed in the locations shown in (e).
\label{fig2}}
\end{center}
\end{figure*}

Fig.~\ref{fig2}(a) shows a schematic illustration of the HMM PhC investigated in this work, including a horizontal dipole \SI{5}{\nm} above it. As a reference in Fig.~\ref{fig2}(b), we plot $\text{PF}_\text{rad}$ spectra of a bare Ag mirror and an unstructured HMM. The Ag mirror is a promising brightening platform, but the dipole must be positioned quite far away from the surface to avoid strong non-radiative emission quenching. Calculated $\text{PF}_\text{rad}$ spectra for a series of HMM PhCs with constant periodicity $a=$ \SI{0.3}{\micro\meter} and various nanohole radii~$r$ is plotted in Fig.~\ref{fig2}(c). We can indeed observe not only the significant increase of $\text{PF}_\text{rad}$ with respect to the unstructured HMM but also broadening of the enhancement region towards infrared wavelengths (cf. Fig.~\ref{fig2}(b)). Furthermore, the HMM PhC still outperforms the bare Ag mirror in the infrared region, and the optimal dipole height of \SI{5}{\nm} is a practically much more viable possibility in terms of fabrication since it can be achieved by direct coating made of quantum dots or by a 2D semiconductor directly on top of the HMM surface. To explain the effect of PhC on HMMs, we recall that high-$k$ modes of all HMMs lie beyond the light line and, although they dominate the enhanced photonic density of states that gives rise to large PFs, they cannot couple to freely propagating radiation and are ultimately lost to absorption. Introducing a periodic nanohole array of period $a$ endows the structure with an additional reciprocal lattice vector, $G = 2\pi/a$, which supplies the missing in-plane momentum required to fold trapped high-$k$ modes back within the light cone, analogous to the operation of a diffraction grating:
\begin{equation}
k_\text{HMM} - G = k_0 \sin\theta,
\label{eq:momentum_matching}
\end{equation}
where $\theta$ is the emission angle of the outcoupled photon relative to the surface normal for an appropriate integer multiple of $G$. Once folded back inside the light cone, these modes can radiate into free space and, in principle, be collected. This momentum-matching mechanism accounts for the enhancement in $\text{PF}_\text{rad}$ observed upon patterning the HMM.

In Fig.\ref{fig2}(c),  the wavelength regions corresponding to the strongest enhancement varied, and the highest $\text{PF}_\text{rad}$ values ($\sim$ 4 -- 5) occured when $r \sim a/2$. The appearance of sharp peaks in $\text{PF}_\text{rad}$ is a result of singularities in the photonic density of states originating from the band structure of the underlying PhC\cite{Joannopoulos2011}. Note that such modest $\text{PF}_\text{rad}$ values in a structured HMM platform are in line with seminal experimental work by Lu et al\cite{Lu2014}. Above, we considered HMM PhCs with etched gratings throughout the full twelve-layer stack (6 pairs of alternating layers). However, such a full etch is associated with an effective loss of HMM volume underneath the quantum emitter. To investigate whether the lost volume contributes positively or negatively to the $\text{PF}_\text{rad}$, we compare an HMM PhC where only 2 pairs of alternating layers were etched with that of a fully-etched HMM PhC of the same $a$ and $r$. As Fig.~\ref{fig2}(d) confirms, the $\text{PF}_\text{rad}$ of the PhC with shallow etching shows lower values, indicating that the high-$k$ modes responsible for the hyperbolic enhancement are distributed throughout the depth of the multilayer stack. Shallow etching perturbs only the near-surface region, leaving some trapped modes unaffected by the grating and thus unable to outcouple, whereas full-depth patterning allows the grating to interact with more high-$k$ modes across the entire stack, maximizing outcoupling efficiency.
This finding indicates the importance of full-height etching of the grating structure.

Spatial distribution of the electric field enhancement above the HMM PhC (for $r=a/2$) is shown in the inset of Fig.~\ref{fig2}(a). The fact that it is concentrated at specific locations rather than distributed uniformly across the unit cell indicates that the $\text{PF}_\text{rad}$ enhancement will originate from localized, grating-coupled modes rather than from a uniform non-local response. Placing the emitter at a location of strong field localization should efficiently couple its near-field to the outcoupled high-$k$ modes, whereas placement away from these regions should largely bypass the grating-mediated outcoupling pathway. We investigate this dependence by evaluating $\text{PF}_\text{rad}$ for six representative in-plane emitter positions spanning all important positions of the dipole relative to the etched HMM grating unit cell (Fig.~\ref{fig2}(e)). The corresponding $\text{PF}_\text{rad}$ spectra at these positions are shown in Fig.~\ref{fig2}(f). $\text{PF}_\text{rad}$ spectra are similar for the centered emitter placements in positions I and II. Interestingly, placing the emitter above a strongly confined region (e.g., positions III and IV) unlocks higher $\text{PF}_\text{rad}$ than positions I and II, but at longer wavelengths still within the expected hyperbolic region. Higher $\text{PF}_\text{rad}$ originates from the extreme confinement of light around PhC, generating very small mode volumes\cite{doi:10.1021/acsphotonics.6b00219,Albrechtsen2022}. The gradual red-shift of the higher $\text{PF}_\text{rad}$ region is observed as the emitter position moves from position II to III and IV. Different emitter positions access different momenta: this spectral red-shift must therefore be related to the dispersion of the allowed optical modes dictated by the underlying band structure, as well as the polariton dispersion in the HMM. Notably, Sekhar \textit{et al.} showed for TiO$_2$/Ag HMMs with a metal fraction of 0.25, that multiple polariton modes appear at the longer near-IR wavelengths as the $k$ vector increases\cite{Shekhar2014}.
We note that precise emitter placement in regions of highest $\text{PF}_\text{rad}$ may be challenging to achieve in practical devices. Nevertheless, this systematic mapping provides clear design guidance for optimizing emitter positioning where such control is achievable. 

\subsection{Tunability of hyperbolicity}

When designing a suitable platform for bright single-photon emitters, external control of brightness is crucial for real-world applications of quantum technologies. To realize external control over radiative Purcell enhancement, we introduce a phase-change material (PCM) as a constituent of the HMM. Antimony trisulfide (Sb$_2$S$_3$) is well known for its phase transition from an amorphous to a crystalline state upon exposure to high temperatures ($\sim$ 543\,K)\cite{Gutierrez2022,Kepic_2024}. The phase change is associated with bandgap narrowing from ~2.1\,eV to ~1.7\,eV, and with a significant change of the material dielectric permittivity in visible and near infrared wavelengths \cite{https://doi.org/10.1002/adfm.201806181}. Light-induced reversible phase change in such materials\cite{https://doi.org/10.1002/adfm.202002447} can be exploited for a tunable HMM device\cite{https://doi.org/10.1002/adom.201800332,https://doi.org/10.1002/adom.201900680}. Notably, Chamoli \emph{et al.} had introduced \cf{Sb2S3}/TiN as a potential platform for phase-change HMM, yet limited their theoretical study to refractive index and variation of PF upon phase change, while experimentally-relevant quantities were not derivable from that work\cite{Chamoli:20}.  

\begin{figure*}[!tb]
\begin{center}
\includegraphics[]{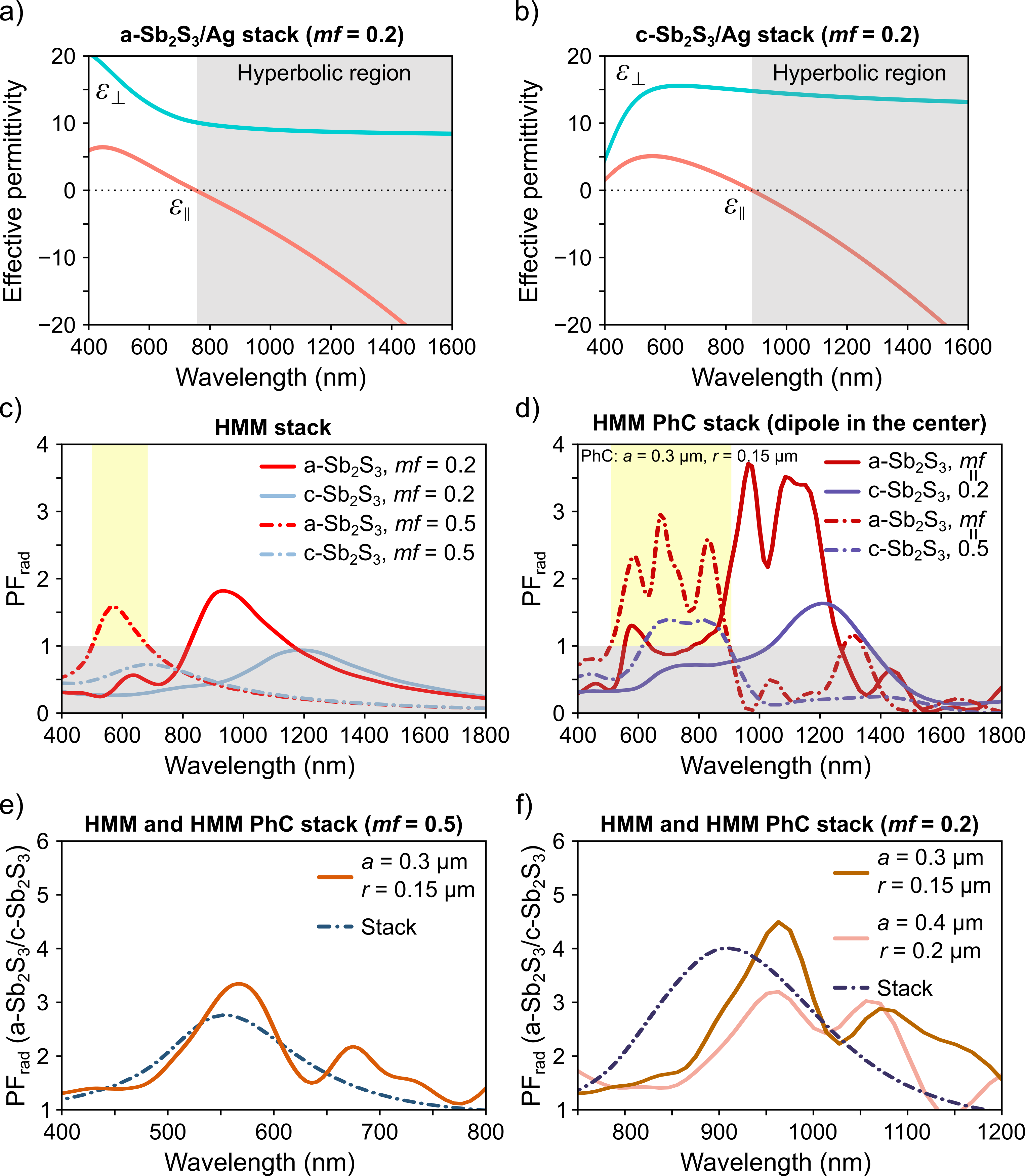}\caption{Tunability of emission enhancement by HMM: In-plane and out-of-plane effective permittivity of a Sb$_2$S$_3/$Ag stack ($f$ = 0.2) in (a) amorphous and (b) crystalline phase of Sb$_2$S$_3$. The hyperbolic region allowing PF$_\text{{rad}} > 1$ is shaded yellow for reference. (c) $\text{PF}_\text{rad}$ of the stack with two different metal fractions (0.2 and 0.5) corresponding to two phases of Sb$_2$S$_3$. (d) $\text{PF}_\text{rad}$ of the nanopatterned HMM PhC with hole periodicity of 0.30\,$\mu$m and radius of 0.15\,$\mu$m. Ratio of the $\text{PF}_\text{rad}$ corresponding to the amorphous to that of the crystalline phase for the HMM stack and HMM PhCs with a metal fraction of (e) 0.5 and (f) 0.2. 
\label{fig3}}
\end{center}
\end{figure*}

To explore the utilization of this material for tunable HMM PhCs, we first consider an HMM stack of Sb$_2$S$_3$ and Ag layers. The effective real parts of the in-plane and out-of-plane permittivities are shown in Fig.~\ref{fig3}(a) and (b) for the amorphous and crystalline phases of the PCM, respectively. A red-shift of the hyperbolic region is noticeable at infrared wavelengths upon switching the material to the crystalline phase. Further, we employed FDTD calculations and plotted the resulting $\text{PF}_\text{rad}$ of the stack at two selected metal fractions (0.2 and 0.5) and at two distinct phases in Fig.~\ref{fig3}(c). Not only can the $\text{PF}_\text{rad}$ be controlled by the phase-change of Sb$_2$S$_3$, but the spectral regions of the hyperbolic enhancements are widely tunable via $f$. Unfortunately, without any extra outcoupling enhancement, low $\text{PF}_\text{rad}$ values ($<2$) are observed. Therefore, we introduced the optimized etched PhC structures into the HMM stack and plotted the corresponding $\text{PF}_\text{rad}$ spectra in Fig.~\ref{fig3}(d). Tunable enhancement of $\text{PF}_\text{rad}$ in the relevant hyperbolic region is noticeable, with $f=0.2$ covering most of the visible range and $f=0.5$ covering most of the near-infrared range. The $\text{PF}_\text{rad}$ tunability is explicitly visualized in Figs. \ref{fig3}(e) and (f), where we plot the relative change of $\text{PF}_\text{rad}$ upon switching between the two phases. A two-fold to five-fold variation of emission brightness is therefore achievable through the PCM-based hyperbolic metamaterials.

\begin{figure*}[tb]
\begin{center}
\includegraphics[width=1.0\linewidth]{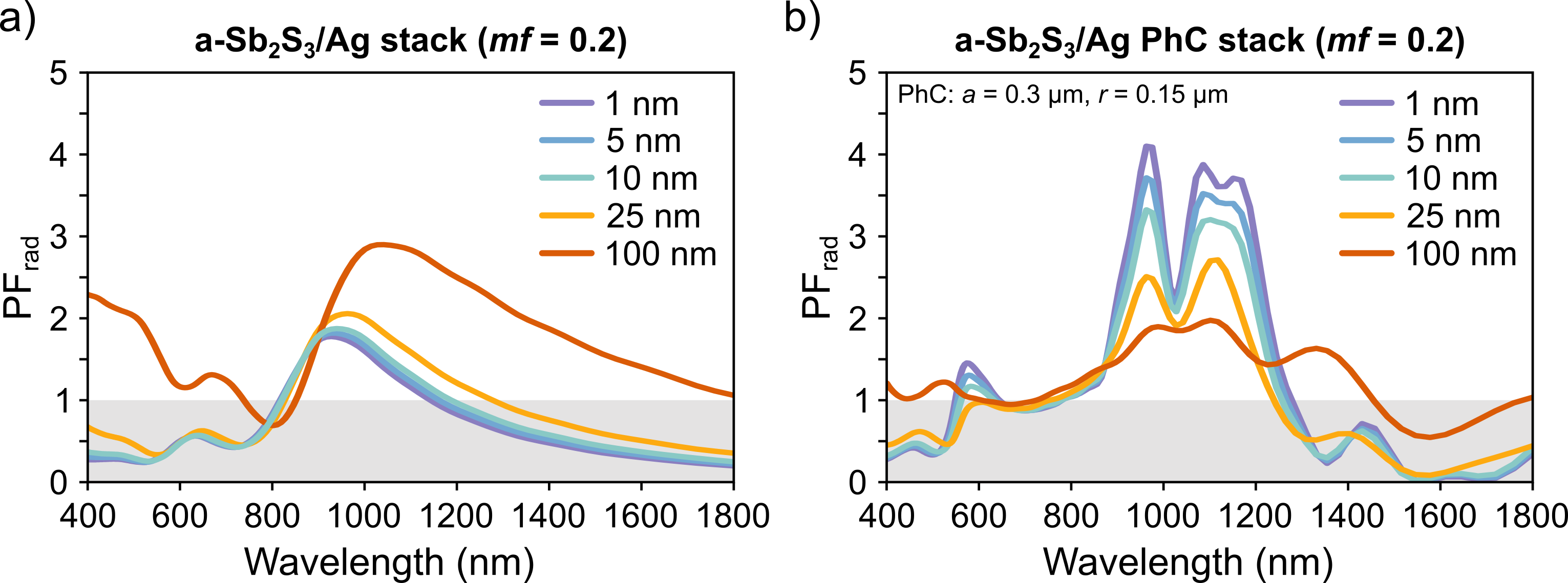}\caption{Height variation of the dipole over an unpatterned HMM stack and a patterned HMM stack: Variation of $\text{PF}_\text{rad}$ of a dipole emitter placed at different heights above the surface of the (a) a-Sb$_2$S$_3$/Ag HMM stack with metal fraction 0.2 and (b) a-Sb$_2$S$_3$/Ag HMM PhC grating with etched nanoholes of periodicity 0.30\,$\mu$m and radius of 0.15\,$\mu$m.
\label{fig4}}
\end{center}
\end{figure*}

In Figures~\ref{fig4}(a) and (b), we thus evaluate spectra of $\text{PF}_\text{rad}$ as a function of emitter heights above the surface of a bare tunable HMM stack and of a tunable HMM PhC, respectively. For the unpatterned Sb$_2$S$_3$/Ag stack ($f = 0.2$), the $\text{PF}_\text{rad}$ spectra remain nearly constant as a function of height from 1\,nm -- 25\,nm. To understand this observation, we need to look into PF and photon out-coupling. The PF is largest for emitters positioned close to the HMM surface and decreases monotonically with increasing height (as demonstrated above in Fig.~\ref{fig1}(e)). In contrast, the non-radiative decay rate, as first demonstrated by Drexhage and co-workers\cite{DREXHAGE1974163}, is strongest at the interface and rapidly diminishes as the emitter is moved away from the surface (see Fig.~\ref{fig1}(e)), before eventually giving rise to the well-known oscillatory behavior associated with interference between the emitter and its image dipole. The interplay between these two opposing trends results in an almost distance-independent $\text{PF}_\text{rad}$ over the hyperbolic spectral window. 
These results indicate that the moderate emission brightening achieved in planar HMM stacks in the hyperbolic region (typically by a factor of 1--2, see Fig.~\ref{fig2}(c)) is largely insensitive to the emitter--surface separation over the range 1\,nm -- 25\,nm. Consequently, similar radiative enhancements are expected for emitters of all sorts of physical dimensions, including quantum dots of varying diameters and atomically thin 2D material emitters.

We next performed the same analysis of $\text{PF}_\text{rad}$ for the nanostructured HMM PhC, placing the emitter at location I of Fig.~\ref{fig2}(e). This location was chosen for its experimental accessibility, as placing an emitter near the center of the structure is more readily achievable in practice than at the more precisely defined edge locations. 
The height-dependent $\text{PF}_\text{rad}$ spectra displayed in Fig.~\ref{fig4}(b) confirm that, as expected, brightness enhancement is substantially stronger in the patterned HMM PhC; this effect is particularly pronounced for emitters located closer to the surface. The nanostructuring clearly suppresses non-radiative decay channels (as opposed to the planar configuration) and enables hyperbolic modes to dominate the spontaneous emission rate enhancement over the spectral range 900\,nm -- 1200\,nm.
The height ($h$) dependence in Fig.~\ref{fig4}(b) reflects the coupling strength between the dipole and the high-$k$ modes: the near field of the emitter contains in-plane wavevectors $k_\parallel > k_0$ which follows evanescent decay as $e^{-\kappa h}$, where $\kappa=\sqrt{k_\parallel^2-k_0^2}$ and $k_0=2\pi/\lambda$. For $k_\parallel \approx 2\pi/a$, decay length is a few tens of nanometers. Consequently, PF$_{rad}$ reduces with increased emitter height from the surface, while the spectral positions of the $\text{PF}_\text{rad}$ peaks remain unchanged. At $h=100$\,nm, where the coupling to the high-$k$ modes has decayed substantially, the patterned stack no longer outperforms the planar one (Fig.~\ref{fig4}(a) and (b)), so the benefit of nanostructuring is restricted to emitters within a few tens of nanometers of the surface.

This finding suggests that two-dimensional emitters, which can be positioned in close proximity to the nanostructured surface, are especially well suited for integration with the HMM PhC platform, enabling efficient enhancement of their emission brightness. 

\begin{figure*}[tb]
    \centering
    \includegraphics[width=1.0\linewidth]{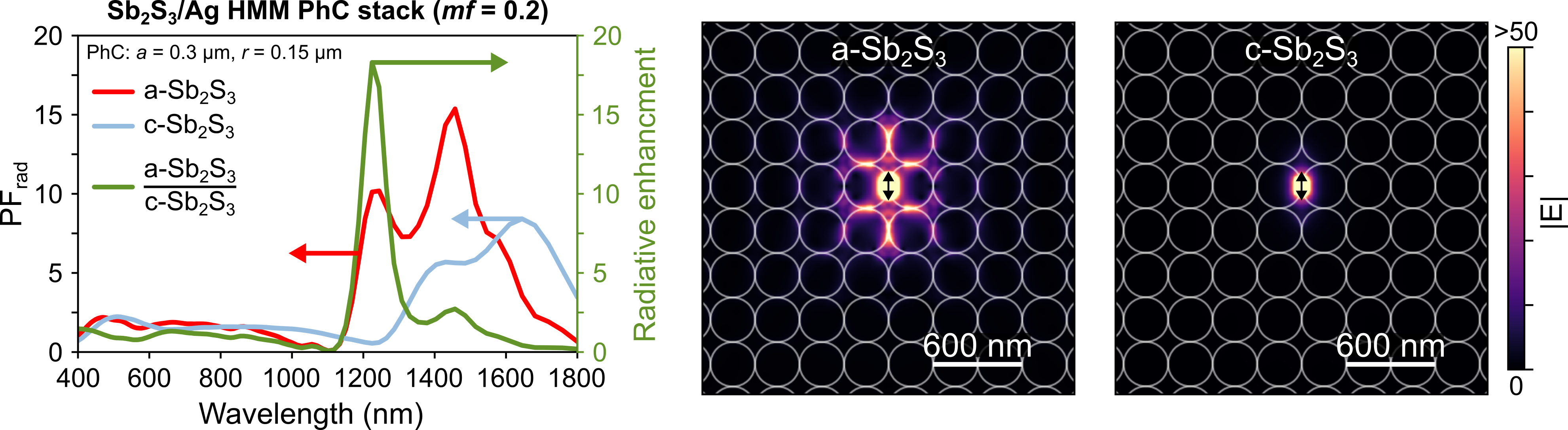}
    
    
    
    \caption{Optimized and tunable single-photon emission configuration: (a) Left axis: $\text{PF}_\text{rad}$ of Sb$_2$S$_3/$Ag HMM PhC (r = a/2 = 0.15\,$\mu$m), Right axis: resulting ratio of the $\text{PF}_\text{rad}$. Simulated electric-field amplitude ($|E|$) corresponding to (b) amorphous and (c) crystalline phase at 1230\,nm.
    \label{fig5}}
\end{figure*}

By combining the insights gained from the in-plane and out-of-plane dependence of the emitter position relative to the HMM PhC, we identify an optimal device configuration for tunable enhancement of light emission: Figure~\ref{fig5} presents the performance of an Sb$_2$S$_3$/Ag HMM ($f=0.2$) PhC with a hole radius of $r=a/2=0.15$\,$\mu$m, where the emitter is positioned \SI{1}{\nm} above the surface and located at the edge between two neighboring patterned holes (position IV in Fig.~\ref{fig2}(e)) for maximal radiative outcoupling. In the amorphous phase, this optimized geometry exhibits an excellent $\text{PF}_\text{rad}$ of approximately 10--15 across the technologically relevant near-infrared spectral range. Upon phase change, the hyperbolic spectral window shifts towards longer wavelengths due to the redshift of the Sb$_2$S$_3$ bandgap. As a result, the $\text{PF}_\text{rad}$ spectrum is also redshifted, enabling a switching contrast exceeding a factor of 15 around the infrared wavelength of 1230\,nm (Figure~\ref{fig5}a). The corresponding near-field electric-field amplitude ($|E|$) distributions at 1230\,nm for the amorphous and crystalline phases are shown in Figs.~\ref{fig5}(b) and \ref{fig5}(c), respectively, illustrating the distinct modal profiles associated with the two material phases.

\section{Conclusion}

In summary, we have investigated the emission properties of quantum emitters in the presence of planar and nanostructured hyperbolic metamaterials and have demonstrated that the conventional PF alone is insufficient for evaluating their performance as photon-brightening platforms. Conventional photonic structures enabling $>100$ PF often possess mere values of EQE $< 0.01$. To address this ambiguity due to disparities in merit functions, we introduced $\text{PF}_\text{rad}$ as a practical figure of merit that directly characterizes the enhancement of quantum emitters. Unlike the conventional PF, which may be dominated by non-radiative decay, the $\text{PF}_\text{rad}$ directly reflects the experimentally observable brightness enhancement and therefore provides a more appropriate metric for evaluating quantum-photonic platforms. Although metal--dielectric HMMs support extremely large photonic densities of states and PFs, much of the enhanced decay occurs through non-radiative channels, resulting in poor photon outcoupling. By incorporating nanopatterning into the form of HMM PhC architecture, high-$k$ hyperbolic modes can be efficiently extracted, leading to significant improvements in radiative emission. We observed an order of magnitude enhancement of the $\text{PF}_\text{rad}$ from 1--2$\times$ in 1D HMMs, and 10--15$\times$ enhancement in structured HMM PhCs. We further showed that integrating a phase-change material Sb$_2$S$_3$ provides active control over the hyperbolic dispersion and enables reversible tuning of the emission spectrum and brightness. Optimizing both the emitter position and the nanostructure geometry yielded $\text{PF}_\text{rad}$ of approximately 10--15$\times$, along with a switching contrast exceeding a factor of 15 in the near-infrared spectral range. These findings highlight the importance of jointly engineering the local density of optical states and photon outcoupling pathways, and establish phase-change HMM photonic crystals as a promising platform for tunable emission brightening and reconfigurable nanophotonic devices.

\begin{acknowledgement}

We thank the Joint Czech-Bavarian Research Projects 2024–2026 program for the research funding via project BTHA-JC-2024-49 (Bavarian-Czech Academic Agency - BTHA) and project LUABA24069 (Ministry of Education, Youth and Sports of the Czech Republic). The Center for Nanoscience (CeNS) of the Ludwig-Maximilians-University Munich (LMU) is also gratefully acknowledged.

\end{acknowledgement}

\subsection{Methods}

\textit{Purcell factor calculation:} PF is defined as the ratio of total power emitted by the emitter in the presence of a target optical environment ($\Gamma$) and that of vacuum ($\Gamma_0$). We use Ansys Lumerical Stack for these calculations. Transfer matrix methods have been used to analytically estimate the PF :
\begin{equation}
\text{PF}= \frac{\Gamma}{\Gamma_0} = \frac{(\Gamma_\text{rad}+\Gamma_\text{non-rad})}{\Gamma_0 }
\end{equation}


\textit{Radiative Purcell factor:} We utilize electromagnetic FDTD simulations using Lumerical to estimate radiative Purcell factors as follows:
$\text{PF}_\text{rad} = \frac{\int _A E^2 dA}{\int _A E_0^2 dA}$, where $E_0$ is the electric field in vacuum and $E = \sqrt{E_x^2+E_y^2+E_z^2}$ is the electric field in the presence of any optical medium. After running convergence tests, a simulation area ($A$) of \SI{6}{\micro\meter} $\times$ \SI{6}{\micro\meter} was chosen. The \SI{6}{\micro\meter} $\times$ \SI{6}{\micro\meter} collection plane positioned \SI{2.5}{\micro\meter} above the surface subtended a solid angle equivalent to a collection cone with numerical aperture NA~$\approx$~0.8, comparable to high-NA objectives typically used in single-photon collection setups, lending the calculated $\text{PF}_\text{rad}$ direct experimental relevance. Perfectly matched layer (PML) boundary conditions were set at all boundaries. The collection plane was placed well-clear of any PML artifacts.

\subsection*{Competing Interests}
The authors declare no competing financial interest.

\subsection*{Data Availability Statement}
The data supporting the findings of this study, including simulation results and analysis scripts used to generate the figures, are available from the corresponding authors upon reasonable request.

\bibliography{References}
\bibstyle{unsrt}
\end{document}